\documentclass[prl,aps,twocolumn,groupedaddress,showpacs]{revtex4}
\begin{document}
\input{psfig.sty}

\title{General Relativistic Decompression of Binary Neutron Stars During 
Dynamic Inspiral}

\author{Mark Miller}
\affiliation{Jet Propulsion Laboratory, California Institute of Technology,
Pasadena, California 91109, USA}

\date{\today}

\begin{abstract}

We investigate the dynamic stability of inspiraling 
neutron stars by performing multiple-orbit 
numerical relativity simulations of the
binary neutron star inspiral process.  
By introducing eccentricities in the orbits of the neutron stars, significant
changes in orbital separation are obtained within orbital timescales.
We find that as the binary system evolves from apastron
to periastron (as the binary separation decreases), the central rest mass
density of each star {\it decreases}, thus stabilizing the
stars against individual prompt collapse.  As the binary system evolves from 
periastron to apastron, the central rest mass density
{\it increases};  the neutron stars re-compress as the binary
separation increases.  

\end{abstract}

\pacs{04.25.Dm, 04.30.Db, 04.40.Dg, 02.60.Cb}

\maketitle

\section{Introduction}

Detailed and accurate models of binary neutron 
star coalescence phenomena, from
quasi-equilibrium orbits through plunge and merger to the subsequent 
formation of the final compact object, will be required in order to 
extract information regarding the structure of neutron stars from 
detected gravitational wave signals.  One detail of some debate over the
past decade has been the 
so-called ``neutron star crushing effect'', reported
in~\cite{Wilson95,WilsonMathewsMarronetti9598,Wilson0005}.
This effect was first reported in~\cite{Wilson95}, 
where binary neutron star
simulations employing a variety of simplifying assumptions
indicated that as
the binary stars spiraled inwards, the 
general relativistic gravitational interaction between
the two stars
caused a destabilization, triggering the collapse of each star to 
individual black holes well {\it before} the plunge and merger phase of
the inspiral.  Since then, studies using several different sets of 
approximations~\cite{Lai96,Wiseman97,Brady97,Baumgarte97,
                     Flanagan98,Thorne:1997kt,Gourgoulhon01,
                     Taniguchi03}
predict the exact opposite, namely that the gravitational
interaction between the two stars act to stabilize each star
as the binary separation decreases during inspiral.  
An error in the formulation used in the original 
``neutron star crushing effect''
studies was pointed out in~\cite{Flanagan99}.  However, subsequent
studies still claim a destabilization 
effect~\cite{Wilson0005}.

In this Article, we report on the first fully dynamical
general relativistic simulation results
aimed at studying this ``neutron star crushing effect'' for
inspiraling binary neutron stars.
In the past several years, 
fully general multiple-orbit binary neutron star simulations 
have been performed in numerical relativity~\cite{Miller04,Marronetti04}.
We use the evolution code 
previously described in~\cite{Miller04} to perform multiple-orbit
simulations of binary neutron stars (for a detailed analysis of the
accuracy of the simulations performed in this 
Article, see~\cite{Miller:2005qu}).
We begin each simulation with
initial data 
corresponding to quasi-equilibrium, circular orbit configurations modified
such that the initial 
orbital angular velocity of the neutron stars is decreased
by varying amounts. 
Each initial data configuration satisfies the initial value constraints
of general relativity.
The simulations resulting from these 
eccentric-orbit initial data sets permit the study, within the
context of full general relativity, of the 
compression/decompression effect by correlating the central 
rest mass density of the neutron stars with the proper separation
of the stars as the evolution progresses through 
several apastron (maximum separation) / periastron 
(minimum separation) points 
during the inspiral.  We find that the stars do, in fact, stabilize 
as the binary separation decreases.  This result is the first fully dynamic
general relativistic demonstration of the decompression of
binary neutron stars;
the typical simplifying 
assumptions made in the treatment of the problem to date,
such as the
truncation of a post-Newtonian expansion, the imposition of 
quasi-equilibrium conditions, or the forcing of the spin states
of the evolution of the neutron stars
(e.g., forcing corotation or irrotation during 
the inspiral), are {\it not} made in the analysis presented here.

\section{Simulation results}
\label{sec:simdetails}

The numerical relativity /
general relativistic hydrodynamics
code used here is described in~\cite{Miller04} and 
analyzed for accuracy in~\cite{Miller:2005qu}.  The grid size used for
simulations is $643 \times 643 \times 325$, 
employing a mirror symmetry about the equatorial plane of the 
neutron stars, $z=0$. 
We choose units such that the gravitational constant $G$ and the speed of
light $c$ are identically $1$.
A spatial discretization of $\Delta x / m = 0.148$ is used, which corresponds
to roughly 40 points across the diameter of 
each neutron star.  The boundary of the domain is thus
eight neutron star diameters away from the center of mass of the 
system.  The physical parameters of three simulations
NS-1, NS-2, and NS-3 are shown in Table~\ref{tab:configs}.
The initial data corresponds to the quasi-equilibrium, corotating, circular
orbit initial data described in~\cite{Miller04}, except that here, 
the initial 
orbital angular velocity parameter $\Omega_0$ has been decreased from
the circular orbit value by a factor of $1\%$, $2\%$, and
$3\%$ for simulation NS-1, NS-2, and NS-3, respectively.  The resulting 
initial data satisfies the Einstein equation initial value constraints, 
and results in binary evolutions whose orbits contain increasing 
amounts of eccentricity.

\begin{table}
\begin{tabular}{|c|c|c|c|c| }  \hline \hline
\hspace{0.0cm} { Config.} \hspace{0.0cm}  &
\hspace{0.0cm} { $m \Omega_0$ } \hspace{0.0cm}  &
\hspace{0.0cm} { $J_0 / m^2$} \hspace{0.0cm}  &
\hspace{0.0cm} { $M_0 / m$} \hspace{0.0cm} &
\hspace{0.0cm} { ${(r_p)}_0 / m$ } \hspace{0.0cm} \\ \hline \hline
{NS-1} &
   { 0.01618} &
   { 1.121} &
   { 1.073} &
   { 18.10} \\ \hline
{NS-2} &
   { 0.01599} &
   { 1.094} &
   { 1.073} &
   { 18.12} \\ \hline
{NS-3} &
   { 0.01580} &
   { 1.067} &
   { 1.073} &
   { 18.14} \\ \hline \hline
\end{tabular}
\caption{The physical parameters of the initial data used for
numerical relativity simulations
NS-1, NS-2, and NS-3.  These initial data sets
correspond to quasi-equilibrium, corotating, circular
orbit initial data with the initial angular velocity $\Omega_0$ reduced from
the circular orbit value by 
$1\%$, $2\%$, and $3\%$, respectively.  
The symbols $J_0$ and $M_0$ are used to denote the initial angular
momentum and total rest mass of the configuration, respectively.
The initial proper
separation of the neutron stars, defined by Eq.~57 in~\cite{Miller04},
is denoted as ${(r_p)}_0$.
A polytropic equation of state with adiabatic index $\Gamma=2$ is used.
The rest mass of each neutron star corresponds to $82\%$ of the
maximum stable TOV rest mass configuration.
All physical quantities in this Article are expressed in units of $m$, which
we define to be twice
the ADM mass of a single
stationary neutron star configuration with rest mass
${M_0}/2$ uniformly rotating with
angular velocity $\Omega_0$.
}
\vspace{0.0cm}
\label{tab:configs}
\end{table}

\begin{figure}
\vspace{0.0cm}
\hspace{0.0cm}
\psfig{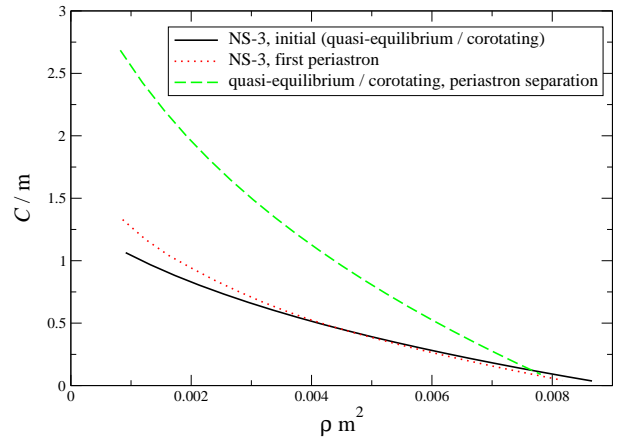}
\caption{The circulation 
$C = \oint_{{\cal{C}}_{\rho}} d\sigma \: h u_{\mu} 
{(\frac {\partial}{\partial \sigma})}^{\mu}$
(h is the relativistic specific enthalpy of the neutron star fluid, 
$u^{\mu}$ is the 4-velocity of the neutron star fluid, and $\sigma$ is
a Lagrange parameter labeling points on the curve)
along constant rest mass density curves ${\cal{C}}_{\rho}$
in the equatorial plane of the neutron stars 
is plotted as a function of rest mass density 
for three different neutron star binary configurations.  The first
(solid) curve shows the circulation of the 
initial corotating, quasi-equilibrium binary configuration
of simulation NS-3, while the second (dotted) curve shows the circulation
of the binary configuration of simulation NS-3 at the time
corresponding to the first periastron point.  For comparison,
the circulation of a corotating, quasi-equilibrium,
equivalent rest mass binary configuration with a proper separation 
equal to that of simulation NS-3 at periastron
is plotted (dashed curve).  Clearly, the spin of the neutron stars
during the numerical evolution NS-3
remains roughly constant.  This is in stark contrast to a 
corotating quasi-equilibrium sequence in which the neutron stars
are artificially ``spun-up'' as the binary separation decreases.
}
\vspace{0.0cm}
\label{fig:circ}
\end{figure}

The initial spin state of each of the neutron stars corresponds
to a corotation with the initial orbital angular velocity.  For binary neutron
stars of $1.4$ solar masses each, this corresponds to a spin period 
of approximately $5$ milliseconds.  However, it is important to note that
during the dynamical evolution, the spins of the neutron stars do
not remain tidally locked to the orbital 
angular velocity (see Fig.~\ref{fig:circ}).  Previous quasi-equilibrium
sequence
inspiral studies have assumed either strictly corotating or
strictly irrotational neutron star spin scenarios;  
in these cases, the spin states
of the neutron stars are fixed to have either corotating
spin or to be irrotational during the inspiral sequence.  
By numerically solving the general coupled 
Einstein/hydrodynamics equations, we are able to study the more 
astrophysically relevant case where the 
non-zero spin of each neutron star actually evolves 
in time via the equations of 
motion.

\begin{figure}
\vspace{0.0cm}
\hspace{0.0cm}
\psfig{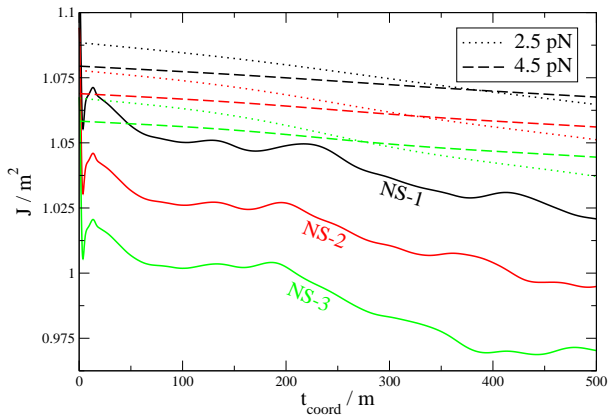}
\caption{The total angular momentum $J$ is plotted for the binary neutron
star numerical relativity simulations NS-1, NS-2, and NS-3 (solid curves).
All simulations correspond to roughly two orbital periods.
For comparison, solutions to the post-Newtonian point particle 
equations of motion are shown, using the same initial coordinate
separation, mass, and initial orbital angular velocity (i.e., $1\%$, $2\%$, 
and $3\%$ below the circular orbit angular velocity) as the 
numerical relativity simulations.  Dotted curves are solutions to the
2.5 post-Newtonian (quadrupole level) equations of motion 
(see, e.g.,~\cite{Pati02}).  Dashed curves are solutions to the 
4.5 post-Newtonian equations of motion, which include 
3.5 pN~\cite{Pati02} and 4.5 pN~\cite{Gopakumar97} radiation reaction
terms, along with the 3.0 pN~\cite{Blanchet03} conservative terms.
}
\vspace{0.0cm}
\label{fig:angmom}
\end{figure}

In Fig.~\ref{fig:angmom}, the angular momentum of the binary system 
simulations is 
plotted as a function of coordinate time;
$t_{coord} = 500 \: m$ corresponds to roughly two orbital periods.   
For comparison, the angular momentum for
solutions to the 2.5 post-Newtonian (dotted curves) and 4.5 post-Newtonian
(dashed curves) point particle equations of motion is plotted along side
the simulation results (solid curves).   The overall rate of decrease
in angular momentum for the numerical relativity simulations is roughly
equivalent, if not slightly more, than predicted by the post-Newtonian
equations of motion.  However, the difference in total angular momentum
between the numerical relativity simulations NS-1, NS-2, and NS-3
(solid curves) is nearly twice as large as the difference
between the post-Newtonian curves, even though the post-Newtonian initial
data is selected using the same method as in the numerical relativity
case, namely, using circular orbit initial data with a decrease in
initial angular velocity 
of $1\%$, $2\%$, and $3\%$ (see Table~\ref{tab:configs}).  This is most
likely due to the nonlinear effects of resolving the constraints of 
general relativity after decreasing the initial angular velocity parameter
$\Omega_0$ when preparing the initial data for the numerical relativity
simulations.

\begin{figure}
\vspace{0.0cm}
\hspace{0.0cm}
\psfig{figure=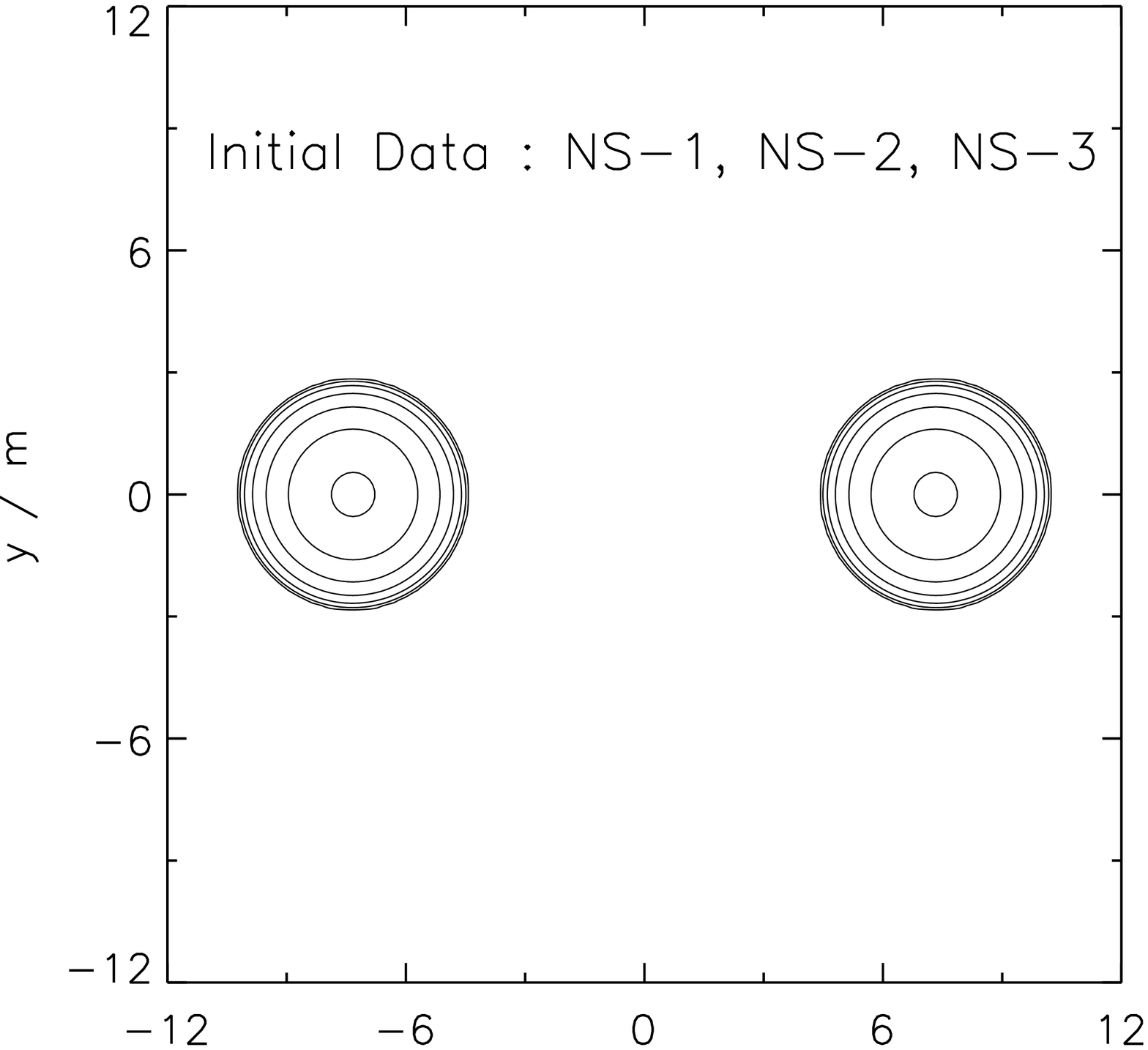,width=4cm}
\psfig{figure=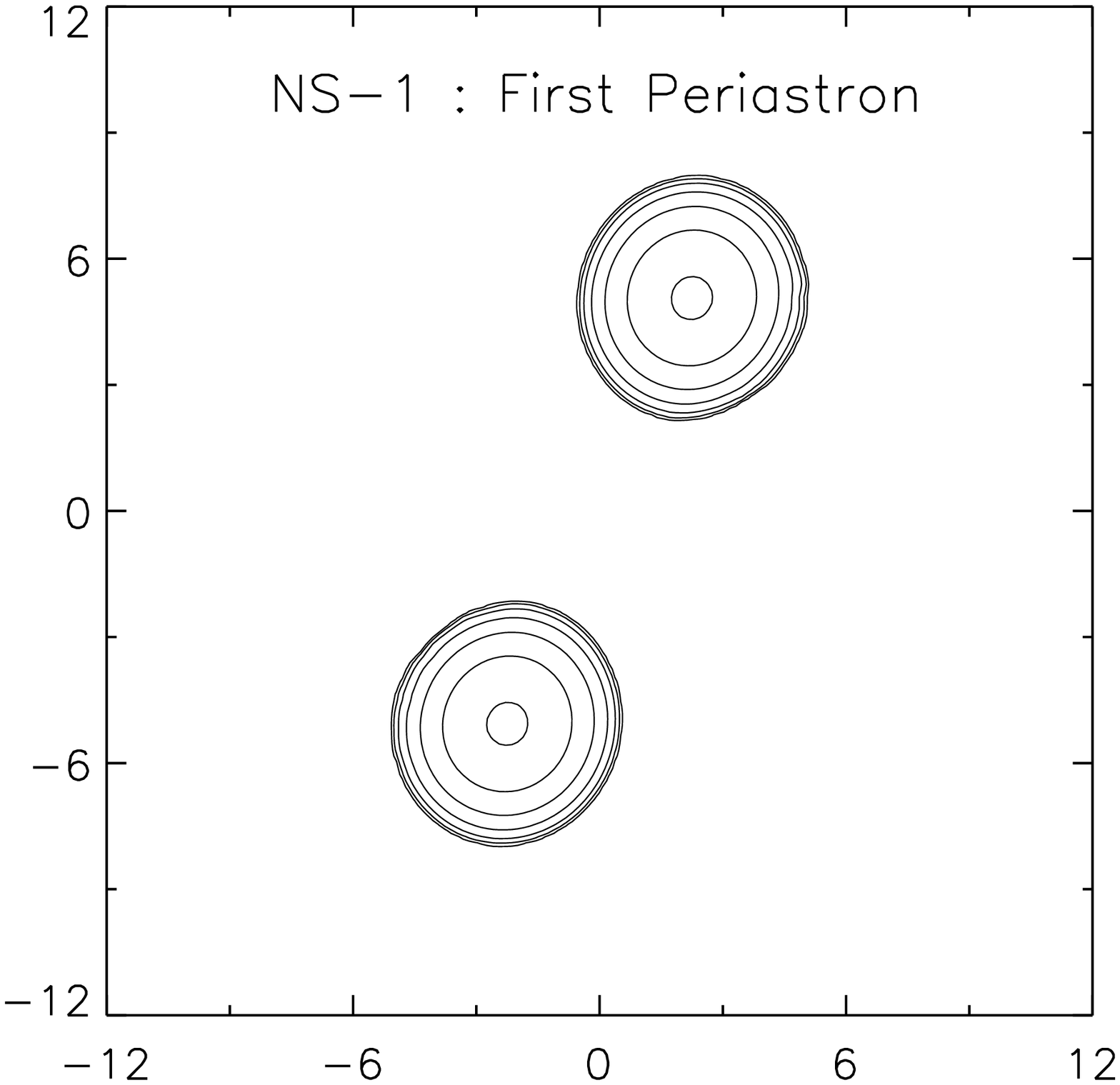,width=4cm}
\psfig{figure=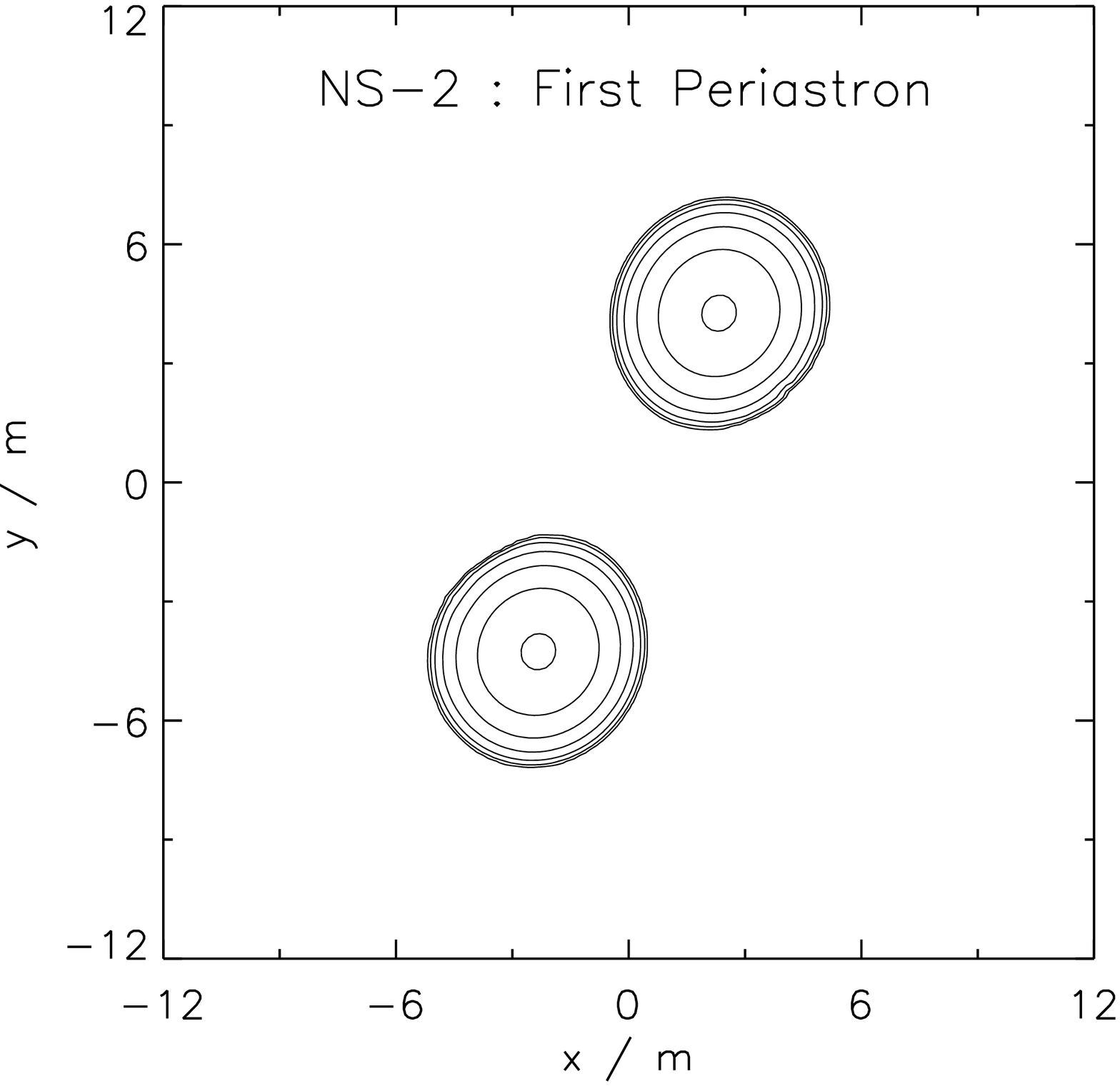,width=4cm}
\psfig{figure=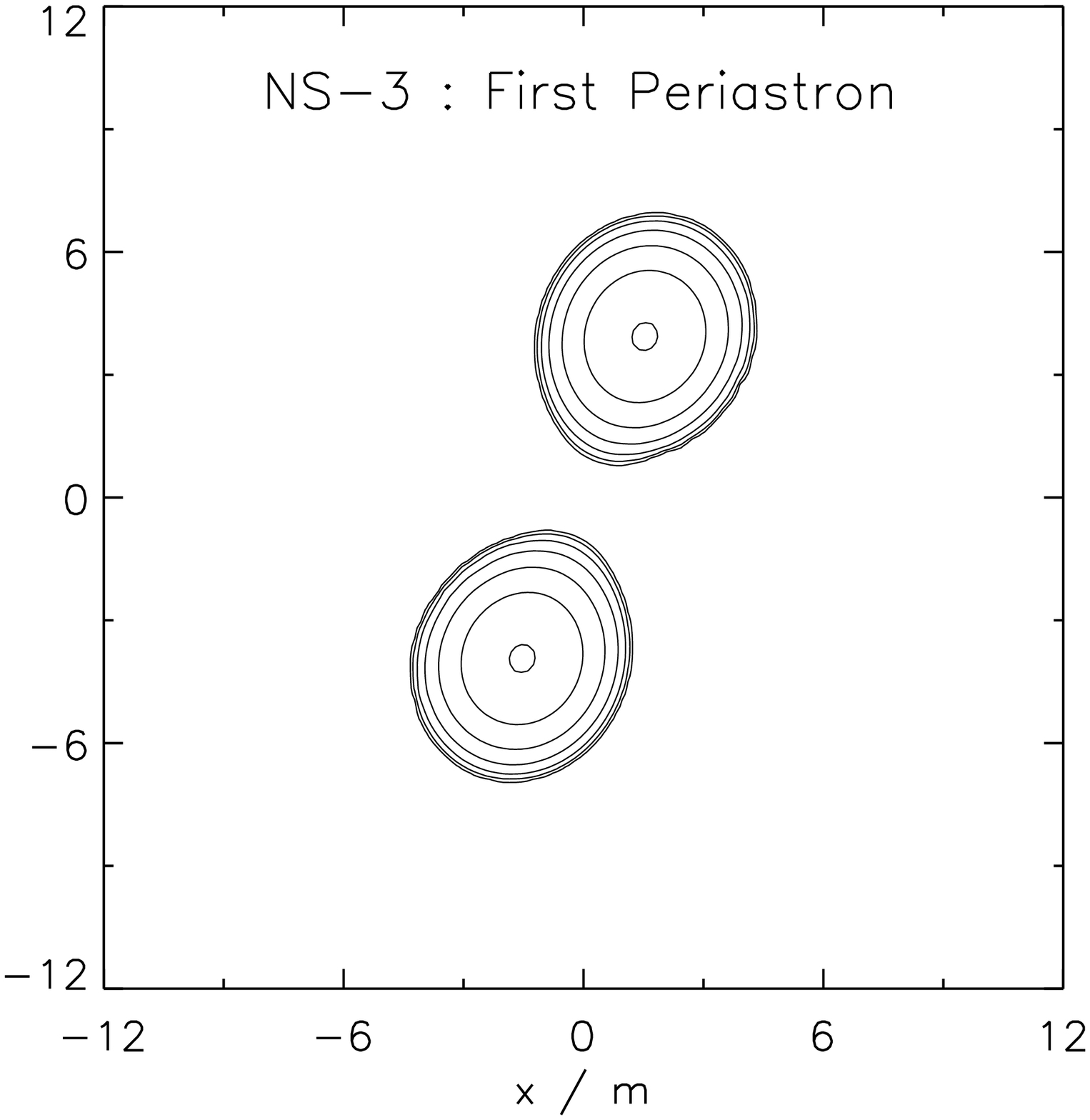,width=4cm}
\caption{Contour plots of the rest mass
density $\rho$ in the $x-y$ equatorial plane of the binary
neutron stars is shown for initial data (upper left) and the first
periastron point for simulations NS-1 (upper right), NS-2 (lower left),
and NS-3 (lower right).  The central contour corresponds to
a value of $\rho = 0.0083 / m^2$.  Six more contours are shown;
each successive contour corresponds to a decrease in rest mass
density by a factor of two.
}
\label{fig:rhocontour}
\end{figure}

Contour plots of the 
rest mass density for the initial data, as well as the first 
periastron points of simulations NS-1, NS-2, and NS-3, are shown in
Fig.~\ref{fig:rhocontour}.  The first periastron points occur just slightly
before the completion of three-quarters of an orbit for all three simulations.

\section{Decompression of binary neutron stars}

\begin{figure}
\vspace{0.0cm}
\hspace{0.0cm}
\psfig{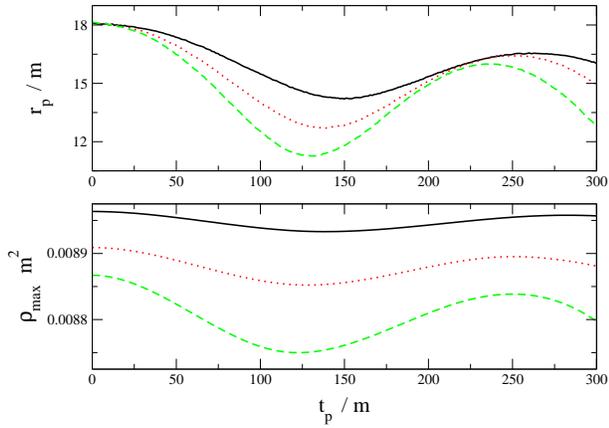}
\caption{Top panel: the proper separation of the binary $r_p$ is plotted
as a function of proper time $t_p$ as measured by observers
at the maximum rest mass density $\rho_{max}$ of the neutron
stars, for numerical relativity
simulations NS-1 (solid curve), NS-2 (dotted curve), and
NS-3 (dashed curve).  Bottom panel: the maximum rest mass density
$\rho_{max}$ of the neutron stars is plotted as a function of
$t_p$ for numerical relativity
simulations NS-1 (solid curve), NS-2 (dotted curve), and
NS-3 (dashed curve).  Oscillations in $\rho_{max}(t_p)$ corresponding
to the fundamental radial oscillation mode of each neutron star
have been filtered out.
}
\vspace{0.0cm}
\label{fig:l_rho_vs_u}
\end{figure}

\begin{figure}
\vspace{0.0cm}
\hspace{0.0cm}
\psfig{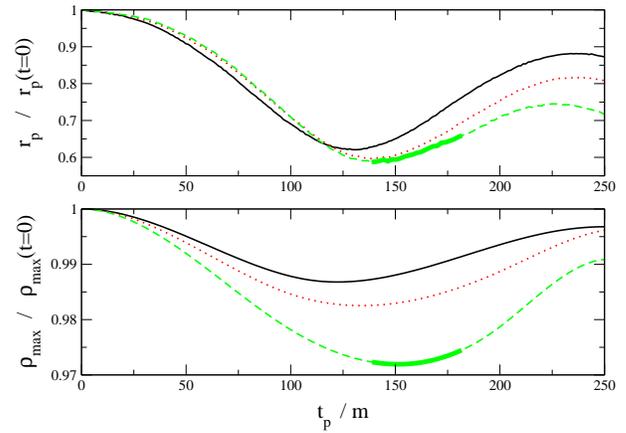}
\caption{The proper separation of the binary (top panel) and maximum rest mass 
density (bottom panel) are plotted as a function of proper time; each 
quantity is normalized by its initial value.  Shown are NS-3 simulation
results (solid curve).  The simulation was repeated for different
resolutions and grid sizes: the dotted curves are results obtained
using $323 \times 323 \times 165$ grid points with 
$\Delta x / m = 0.211$;  the dashed curves are results obtained using
$195 \times 195 \times 101$ grid points with
$\Delta x / m = 0.247$ (the thick, solid sections of these 
low-resolution curves represent when the neutron stars are in contact
with each other).
The Richardson extrapolation of the 
correlation ${\sf Cor}(r_p,{\rho}_{max})$ between
the proper binary separation and the maximum rest mass density
is $1.00$.
}
\vspace{0.0cm}
\label{fig:convergence}
\end{figure}

In Fig.~\ref{fig:l_rho_vs_u}, we plot the proper binary
separation $r_p$ (defined to be the spatial geodesic distance between
the maximum rest mass density point of each neutron star, see
Eq.~57 in~\cite{Miller04})
as a function of the
proper time $t_{p}$ measured by observers located at the maximum
rest mass density points of the neutron stars.  An
evolved proper time of $t_{p} = 300 \: m$ corresponds to roughly 
two orbital periods.
Also in Fig.~\ref{fig:l_rho_vs_u} is a plot
of the maximum rest mass density $\rho_{max}$ 
of the neutron stars as a function
of proper time $t_{p}$.  Oscillations in $\rho_{max}(t_{p})$
corresponding to the fundamental radial oscillation mode of each 
neutron star have been filtered out (these oscillations have a proper
period of $T_{p} = 35.0 \: m$).
Fig.~\ref{fig:l_rho_vs_u} displays a clear correlation
between the maximum rest mass density of the stars and the proper 
separation of the stars;  smaller proper separation $r_p$ 
corresponds to
smaller maximum rest mass density $\rho_{max}$.  
More quantitatively, the correlation ${\sf Cor}(r_p,{\rho}_{max})$ between
the proper binary separation and the maximum rest mass density
is computed to be
\begin{equation}
{\sf Cor}(r_p,{\rho}_{max}) > 0.95
\label{eq:cor}
\end{equation}
for all simulations.  Convergence tests on simulation NS-3 yields 
a Richardson extrapolation value of ${\sf Cor}(r_p,{\rho}_{max}) = 1.00$
(see Fig.~\ref{fig:convergence}).

Fig.~\ref{fig:l_rho_vs_u} also shows that the correlation 
between orbital separation and maximum rest mass density exists
over gravitational radiation
reaction timescales.  
This is evidenced by comparing both the proper separation of the 
binary and the maximum rest mass density at successive apastron (local
maximum in separation) points in Fig.~\ref{fig:l_rho_vs_u}.  First,
we see that the proper separation $r_p$ at the second
apastron point (which occurs at roughly $t_p = 250 \: m$
for all three simulations) is approximately $10\%$ smaller than the
proper separation at the
first apastron point (which occurs at
the initial time $t_{p} = 0$).  Gravitational radiation
emission drains the binding energy of the binary, causing an
overall decrease in separation from one apastron point to
the next.
We also see from Fig.~\ref{fig:l_rho_vs_u} that the maximum rest mass
density of the neutron stars is smaller at the second apastron point 
as compared to the first apastron point.  Thus, these
simulations also demonstrate the correlation between orbital 
separation and maximum rest mass density on gravitational radiation
reaction timescales.

In~\cite{Flanagan98},
a post-Newtonian matched asymptotic expansion technique
is used to obtain an expression for the fractional change
in the central density ${\rho}_c$ 
of inspiraling binary neutron stars, perturbative in
powers of the
``tidal expansion parameter'' $\alpha \equiv R/r$ where $R$ is the
stellar radius and $r$ is the binary separation, such that
\begin{equation}
\frac {\delta \rho_c}{\rho_c} \propto {\alpha}^6
\label{eq:alphascale}
\end{equation}
as $\alpha \rightarrow 0$.  That is, when the fractional change in the
central rest mass density of the stars is expanded about 
$\alpha = 0$ (i.e. infinite binary separation) in powers of
$\alpha$, post-Newtonian theory predicts the first non-zero term to
be $\alpha^6$.  We note that the limited range of binary separations, 
along with the large values of $\alpha$ obtained in our simulations
($\alpha = R_p / r_p \approx 0.4$, where $R_p$ is the proper stellar
radius computed from the proper stellar volume $V_p$ according
to $R_p = {(3 V_p / (4 \pi))}^{1/3}$), prevent us from accurately
determining the power of the first non-zero term in 
an expansion of the fractional change in the central 
rest mass density as in, e.g., Eq.~\ref{eq:alphascale}.
Simulations involving much smaller values of $\alpha$ (larger
binary separations, which imply longer orbital periods), which
will only be capable with adaptive mesh refinement technology
and/or higher order methods, will be required to verify 
the post-Newtonian prediction Eq.~\ref{eq:alphascale}.

\section{Acknowledgment}

It is a pleasure to thank my colleagues
at the Jet Propulsion 
Laboratory 
for many useful 
discussions and suggestions.
Financial support for this research has been
provided by the
Jet Propulsion Laboratory under contract with the
National Aeronautics and Space Administration.
Computational resource support has been provided by the
JPL Institutional Computing and Information Services, 
the NASA Directorates of Aeronautics Research, Science, Exploration
Systems, and Space Operations, and 
NSF NRAC project MCA02N022.



\end{document}